\documentclass[a4paper,10pt]{article}
\usepackage{graphicx,amssymb,epsfig,epsf}
\newcommand{\newc}{\newcommand}

\def\Ord{\lower .7ex\hbox{$\;\stackrel{\textstyle <}{\sim}\;$}}
\def\OOrd{\lower .7ex\hbox{$\;\stackrel{\textstyle >}{\sim}\;$}}
\newc{\order}{{\cal O}}

\def\lum             {{\cal L}}

\newc{\be}{\begin{equation}}
\newc{\ee}{\end{equation}}
\newc{\br}{\begin{eqnarray}}
\newc{\er}{\end{eqnarray}}
\newc{\ba}{\begin{array}}
\newc{\ea}{\end{array}}
\newc{\bi}{\begin{itemize}}
\newc{\ei}{\end{itemize}}
\newc{\bn}{\begin{enumerate}}
\newc{\en}{\end{enumerate}}
\newc{\bc}{\begin{center}}
\newc{\ec}{\end{center}}
\newc{\ul}{\underline}
\newc{\ol}{\overline}
\newc{\ra}{\rightarrow}
\newc{\lra}{\longrightarrow}
\newc{\wt}{\widetilde}
\newc{\til}{\tilde}

\newc{\wh}{\widehat}
\newc{\ti}{\times}
\newc{\Dir}{\kern -6.4pt\Big{/}}
\newc{\Dirin}{\kern -10.4pt\Big{/}\kern 4.4pt}
\newc{\DDir}{\kern -10.6pt\Big{/}}
\newc{\DGir}{\kern -6.0pt\Big{/}}
\newc{\sig}{\sigma}
\newc{\sigmalstop}{\sig_{\lstoppair}}
\newc{\Sig}{\Sigma}  %%\def\Sig{\Sigma} this is the def style
\newc{\del}{\delta}
\newc{\Del}{\Delta}
\newc{\lam}{\lambda}
\newc{\Lam}{\Lambda}
\newc{\gam}{\gamma}
\newc{\Gam}{\Gamma}
\newc{\eps}{\epsilon}
\newc{\Eps}{\Epsilon}
\newc{\kap}{\kappa}
\newc{\Kap}{\Kappa}
\newc{\modulus}[1]{\left| #1 \right|}
\newc{\eq}[1]{(\ref{eq:#1})}
\newc{\eqs}[2]{(\ref{eq:#1},\ref{eq:#2})}
\newc{\etal}{{\it et al.}\ }
\newc{\ibid}{{\it ibid}.}
\newc{\ibidem}{{\it ibidem}.}
\newc{\eg}{{\it e.g.}\ }
\newc{\ie}{{\it i.e.}\ }

\newc{\nonum}{\nonumber}
\newc{\lab}[1]{\label{eq:#1}}
\newc{\dpr}[2]{({#1}\cdot{#2})}
\newc{\lt}{\stackrel{<}}
\newc{\gt}{\stackrel{>}}
\newc{\lsimeq}{\stackrel{<}{\sim}}
\newc{\gsimeq}{\stackrel{>}{\sim}}
\def\lsim{\buildrel{\scriptscriptstyle <}\over{\scriptscriptstyle\sim}}
\def\gsim{\buildrel{\scriptscriptstyle >}\over{\scriptscriptstyle\sim}}
\def\lapp{\mathrel{\rlap{\raise.5ex\hbox{$<$}}
                    {\lower.5ex\hbox{$\sim$}}}}
\def\gapp{\mathrel{\rlap{\raise.5ex\hbox{$>$}}
                    {\lower.5ex\hbox{$\sim$}}}}
\newc{\half}{\frac{1}{2}}

\newc{\bQ}{\ol{Q}}
\newc{\dota}{\dot{\alpha }}
\newc{\dotb}{\dot{\beta }}
\newc{\dotd}{\dot{\delta }}
\newc{\nindnt}{\noindent}

\newc{\matth}{\mathsurround=0pt}
\def\ML{\ifmmode{{\mathaccent"7E M}_L}
             \else{${\mathaccent"7E M}_L$}\fi}
\def\MR{\ifmmode{{\mathaccent"7E M}_R}
             \else{${\mathaccent"7E M}_R$}\fi}

\newc{\mr}{\mathrm}

\newc{\siminf}{\mbox{$_{\sim}$ {\small {\hspace{-1.em}{$<$}}}    }}
\newc{\simsup}{\mbox{$_{\sim}$ {\small {\hspace{-1.em}{$>$}}}    }}

\newc {\Zboson}{{\mathrm Z}^{0}}
\newc{\thetaw}{\theta_W}
\newc{\mbot}{{m_b}}
\newc{\mtop}{{m_t}}
\newc{\sm}{${\cal {SM}}$}
\newc{\as}{\alpha_s}
\newc{\aem}{\alpha_{em}}

\newc{\ppbar}{\mbox{$p\ol{p}$}}
\newc{\bbbar}{\mbox{$b\ol{b}$}}
\newc{\ccbar}{\mbox{$c\ol{c}$}}
\newc{\ttbar}{\mbox{$t\ol{t}$}}
\newc{\eebar}{\mbox{$e\ol{e}$}}
\newc{\zzero}{\mbox{$Z^0$}}

\newc{\wplus}{\mbox{$W^+$}}
\newc{\wminus}{\mbox{$W^-$}}
\newc{\ellp}{\ell^+}
\newc{\ellm}{\ell^-}
\newc{\elp}{\mbox{$e^+$}}
\newc{\elm}{\mbox{$e^-$}}
\newc{\elpm}{\mbox{$e^{\pm}$}}
\newc{\qbar}     {\mbox{$\ol{q}$}}
\newc{\Ebar}{{\bar E}}
\newc{\Dbar}{{\bar D}}
\newc{\Ubar}{{\bar U}}
\newc{\susy}{{{SUSY}}}
\newc{\msusy}{{{M_{SUSY}}}}
\def\photino{\ifmmode{\mathaccent"7E \gam}\else{$\mathaccent"7E \gam$}\fi}
\def\taugluino{\ifmmode{\tau_{\mathaccent"7E g}}
             \else{$\tau_{\mathaccent"7E g}$}\fi}
\def\mphotino{\ifmmode{m_{\mathaccent"7E \gam}}
             \else{$m_{\mathaccent"7E \gam}$}\fi}
\newc{\gl}   {\mbox{$\wt{g}$}}
\newc{\mgl}  {\mbox{$m_{\gl}$}}
\def \chone {{\wt\chi_1}}
\def \ch2p {{\wt\chi_2^+}}

\def \ch2m {{\wt\chi_2^-}}
\def \chplus {{\wt\chi^+}}
\def \chminus {{\wt\chi^-}}

\def \chonepm{{\wt\chi_1}^{\pm}}

\def \mchonepm{m_{\chonepm}}

\def \chtwopm{{\wt\chi_2}^{\pm}}
\def \mchtwopm{m_{\chtwopm}}
\newc{\dmchi}{\Delta m_{\wt\chi}}

\def \lspone{\wt\chi_1^0}
\def \mlspone{m_{\lspone}}
\def \lsptwo{\wt\chi_2^0}
\def \mlsptwo{m_{\lsptwo}}
\def \lspthree{\wt\chi_3^0}
\def \mlspthree{m_{\lspthree}}
\def \lspfour{\wt\chi_4^0}
\def \mlspfour{m_{\lspfour}}

\newc{\sele}{\wt{\mathrm e}}
\newc{\sell}{\wt{\ell}}

\def \stauone{\wt\tau}
\def \mstauone{m_{\stauone}}

\newc{\snue}     {\mbox{$ \wt{\nu_e}$}}
\newc{\smu}{\wt{\mu}}
\newc{\stau}{\wt{\tau}}
\newc {\nuL} {\wt{\nu}_L}
\newc {\nuR} {\wt{\nu}_R}
\newc {\snub} {\bar{\wt{\nu}}}
\newc {\eL} {\wt{e}_L}
\newc {\eR} {\wt{e}_R}
\def \stau{\wt\tau}

\def \sq{\wt{q}}

\newc{\msqot}  {\mbox{$m_(\sq_{1,2} )$}}
\newc{\sqbar}    {\mbox{$\bar{\wt{q}}$}}
\newc{\ssb}      {\mbox{$\squark\ol{\squark}$}}
\newc {\qL} {\wt{q}_L}
\newc {\qR} {\wt{q}_R}
\newc {\uL} {\wt{u}_L}
\newc {\uR} {\wt{u}_R}
\def \ul{\wt{u}_L}
\def \mul{m_{\ul}}
\newc {\dL} {\wt{d}_L}
\newc {\dR} {\wt{d}_R}
\newc {\cL} {\wt{c}_L}
\newc {\cR} {\wt{c}_R}
\newc {\sL} {\wt{s}_L}
\newc {\sR} {\wt{s}_R}
\newc {\tL} {\wt{t}_L}
\newc {\tR} {\wt{t}_R}
\newc {\stb} {\ol{\wt{t}}_1}
\newc {\sbot} {\wt{b}_1}
\newc {\msbot} {m_{\sbot}}
\newc {\sbotb} {\ol{\wt{b}}_1}
\newc {\bL} {\wt{b}_L}
\newc {\bR} {\wt{b}_R}
\def \mul{m_{\wt{u}_L}}
\def \mur{m_{\wt{u}_R}}
\def \mdl{m_{\wt{d}_L}}
\def \mdr{m_{\wt{d}_R}}

\newc{\csquark}  {\mbox{$\wt{c}$}}
\newc{\csquarkl} {\mbox{$\wt{c}_L$}}
\newc{\mcsl}     {\mbox{$m(\csquarkl)$}}

\newc {\stopl}         {\wt{\mathrm{t}}_{\mathrm L}}
\newc {\stopr}         {\wt{\mathrm{t}}_{\mathrm R}}
\newc {\stoppair}      {\wt{\mathrm{t}}_{1}
\bar{\wt{\mathrm{t}}}_{1}}
\def \lstop{\wt{t}_{1}}

\def \lstoppair{\lstop\lstop^*}

\newc{\tsquark}  {\mbox{$\wt{t}$}}
\newc{\ttb}      {\mbox{$\tsquark\ol{\tsquark}$}}
\newc{\ttbone}   {\mbox{$\tsquark_1\ol{\tsquark}_1$}}
\newc{\mix}{\theta_{\wt t}}
\newc{\cost}{\cos{\theta_{\wt t}}}
\newc{\sint}{\sin{\theta_{\wt t}}}
\newc{\costloop}{\cos{\theta_{\wt t_{loop}}}}
\newc{\mixsbot}{\theta_{\wt b}}

\newc{\tb}{\tan\beta}
\newc{\cb}{\cot\beta}
\newc{\vev}[1]{{\left\langle #1\right\rangle}}

\newc{\mhalf}{m_{1/2}}
\newc{\mzero} {\mbox{$m_0$}}
\newc{\azero} {\mbox{$A_0$}}

\newc{\lb}{\lam}
\newc{\lbp}{\lam^{\prime}}
\newc{\lbpp}{\lam^{\prime\prime}}
\newc{\rpv}{{\not \!\! R_p}}
\newc{\rpvm}{{\not  R_p}}
\newc{\rp}{R_{p}}
\newc{\rpmssm}{{RPC MSSM}}
\newc{\rpvmssm}{{RPV MSSM}}

\newc{\sbyb}{S/$\sqrt B$}
\newc{\pelp}{\mbox{$e^+$}}
\newc{\pelm}{\mbox{$e^-$}}
\newc{\pelpm}{\mbox{$e^{\pm}$}}
\newc{\epem}{\mbox{$e^+e^-$}}
\newc{\lplm}{\mbox{$\ell^+\ell^-$}}

\def\Ecm{\ifmmode{E_{\mathrm{cm}}}\else{$E_{\mathrm{cm}}$}\fi}
\newc{\rts}{\sqrt{s}}
\newc{\rtshat}{\sqrt{\hat s}}
\newc{\gev}{\,GeV}
\newc{\mev}{~{\rm MeV}}
\newc{\tev}  {\mbox{$\;{\rm TeV}$}}
\newc{\gevc} {\mbox{$\;{\rm GeV}/c$}}
\newc{\gevcc}{\mbox{$\;{\rm GeV}/c^2$}}
\newc{\intlum}{\mbox{${ \int {\cal L} \; dt}$}}
\newc{\call}{{\cal L}}
\newc{\ptmiss}{/ \hskip-7pt p_T}

\def \etslash{\not \! E_T }

\newc{\PT}{\mbox{$p_T$}}
\newc{\ET}{\mbox{$E_T$}}
\newc{\dedx}{\mbox{${\rm d}E/{\rm d}x$}}
\newc{\ifb}{\mbox{${\rm fb}^{-1}$}}
\newc{\ipb}{\mbox{${\rm pb}^{-1}$}}
\newc{\pb}{~{\rm pb}}
\newc{\fb}{~{\rm fb}}
\newc{\ycut}{y_{\mathrm{cut}}}
\newc{\chis}{\mbox{$\chi^{2}$}}

\def \jet(s){\emph{jet(s) }}

\newc{\mpl}{M_{\rm Pl}}
\newc{\mgut}{M_{GUT}}
\newc{\mw}{M_{W}}
\newc{\mweak}{M_{weak}}
\newc{\mz}{M_{Z}}

\newc{\OPALColl}   {OPAL Collaboration}
\newc{\ALEPHColl}  {ALEPH Collaboration}
\newc{\DELPHIColl} {DELPHI Collaboration}
\newc{\XLColl}     {L3 Collaboration}
\newc{\JADEColl}   {JADE Collaboration}
\newc{\CDFColl}    {CDF Collaboration}
\newc{\DXColl}     {D0 Collaboration}
\newc{\HXColl}     {H1 Collaboration}
\newc{\ZEUSColl}   {ZEUS Collaboration}
\newc{\LEPColl}    {LEP Collaboration}
\newc{\ATLASColl}  {ATLAS Collaboration}
\newc{\CMSColl}    {CMS Collaboration}
\newc{\UAColl}    {UA Collaboration}
\newc{\KAMLANDColl}{KamLAND Collaboration}
\newc{\IMBColl}    {IMB Collaboration}
\newc{\KAMIOColl}  {Kamiokande Collaboration}
\newc{\SKAMIOColl} {Super-Kamiokande Collaboration}
\newc{\SUDANTColl} {Soudan-2 Collaboration}
\newc{\MACROColl}  {MACRO Collaboration}
\newc{\GALLEXColl} {GALLEX Collaboration}
\newc{\GNOColl}    {GNO Collaboration}
\newc{\SAGEColl}  {SAGE Collaboration}
\newc{\SNOColl}  {SNO Collaboration}
\newc{\CHOOZColl}  {CHOOZ Collaboration}
\newc{\PDGColl}  {Particle Data Group Collaboration}

\def\issue(#1,#2,#3){{\bf #1}, #2 (#3)}%AIP format!Vol,page(Year)
\def\ASTR(#1,#2,#3){Astropart.\ Phys. \issue(#1,#2,#3)}
\def\AJ(#1,#2,#3){Astrophysical.\ Jour. \issue(#1,#2,#3)}
\def\AJS(#1,#2,#3){Astrophys.\ J.\ Suppl. \issue(#1,#2,#3)}
\def\APP(#1,#2,#3){Acta.\ Phys.\ Pol. \issue(#1,#2,#3)}
\def\JCAP(#1,#2,#3){Journal\ XX. \issue(#1,#2,#3)} %spdas
\def\SC(#1,#2,#3){Science \issue(#1,#2,#3)}
\def\PRD(#1,#2,#3){Phys.\ Rev.\ D \issue(#1,#2,#3)}
\def\PR(#1,#2,#3){Phys.\ Rev.\ \issue(#1,#2,#3)} % spdas check 
\def\PRC(#1,#2,#3){Phys.\ Rev.\ C \issue(#1,#2,#3)}
\def\NPB(#1,#2,#3){Nucl.\ Phys.\ B \issue(#1,#2,#3)}
\def\NPPS(#1,#2,#3){Nucl.\ Phys.\ Proc. \ Suppl \issue(#1,#2,#3)}
\def\NJP(#1,#2,#3){New.\ J.\ Phys. \issue(#1,#2,#3)}
\def\JP(#1,#2,#3){J.\ Phys.\issue(#1,#2,#3)}
\def\PL(#1,#2,#3){Phys.\ Lett. \issue(#1,#2,#3)}
\def\PLB(#1,#2,#3){Phys.\ Lett.\ B  \issue(#1,#2,#3)}
\def\ZP(#1,#2,#3){Z.\ Phys. \issue(#1,#2,#3)}
\def\ZPC(#1,#2,#3){Z.\ Phys.\ C  \issue(#1,#2,#3)}
\def\PREP(#1,#2,#3){Phys.\ Rep. \issue(#1,#2,#3)}
\def\PRL(#1,#2,#3){Phys.\ Rev.\ Lett. \issue(#1,#2,#3)}
\def\MPL(#1,#2,#3){Mod.\ Phys.\ Lett. \issue(#1,#2,#3)}
\def\RMP(#1,#2,#3){Rev.\ Mod.\ Phys. \issue(#1,#2,#3)}
\def\SJNP(#1,#2,#3){Sov.\ J.\ Nucl.\ Phys. \issue(#1,#2,#3)}
\def\CPC(#1,#2,#3){Comp.\ Phys.\ Comm. \issue(#1,#2,#3)}
\def\IJMPA(#1,#2,#3){Int.\ J.\ Mod. \ Phys.\ A \issue(#1,#2,#3)}
\def\MPLA(#1,#2,#3){Mod.\ Phys.\ Lett.\ A \issue(#1,#2,#3)}
\def\PTP(#1,#2,#3){Prog.\ Theor.\ Phys. \issue(#1,#2,#3)}
\def\RMP(#1,#2,#3){Rev.\ Mod.\ Phys. \issue(#1,#2,#3)}
\def\NIMA(#1,#2,#3){Nucl.\ Instrum.\ Methods \ A \issue(#1,#2,#3)}
\def\JHEP(#1,#2,#3){J.\ High\ Energy\ Phys. \issue(#1,#2,#3)}
\def\EPJC(#1,#2,#3){Eur.\ Phys.\ J.\ C \issue(#1,#2,#3)}
\def\RPP (#1,#2,#3){Rept.\ Prog.\ Phys. \issue(#1,#2,#3)}
\def\PPNP(#1,#2,#3){ Prog.\ Part.\ Nucl.\ Phys. \issue(#1,#2,#3)}
\def\JPG(#1,#2,#3){J.\ Phys.\ G  \issue(#1,#2,#3)} 
\newc{\PRDR}[3]{{Phys. Rev. D} {\bf #1}, Rapid  Communications, #2 (#3)}
\def \selecl{{\wt e_L}}
\def \mselecl{m_{\selecl}}
\def \selecr{{\wt e_R}}
\def \mselecr{m_{\selecr}}
\def \stauone{{\wt \tau_1}}
\def \mstauone{m_{\stauone}}
\def \stautwo{{\wt \tau_2}}
\def \mstautwo{m_{\stautwo}}
\def \staunu{{\wt \nu_\tau}}
\def \mstaunu{{m_{\staunu}}}

\def \selecnu{{\wt \nu_e}}
\def \mselecnu{{m_{\selecnu}}}
\def \mbone{m_{\wt{b}_1}}
\def \mbtwo{m_{\wt{b}_2}}
\def \mtone{m_{\wt{t}_1}}
\def \mttwo{m_{\wt{t}_2}}

\def \mhiggsone{m_h}
\def \mhiggstwo{m_H}
\def \mhiggsthree{m_A}
\def \mchhiggs{m_{H^{\pm}}}

\begin{document}
%\linenumbers
\tolerance=100000
\thispagestyle{empty}
\setcounter{page}{0}

%-----------------------------------------

%\vspace*{\fill}
%\vspace{-1.5in}
\begin{flushright}
{\tt TIFR/TH/11-16}\\
{\tt IISER/HEP/02/11}\\
\end{flushright}
\begin{center}
{\Large \bf Revealing the footprints of  
squark gluino production through Higgs search experiments at the Large 
Hadron Collider
at 7 TeV and 14 TeV}
 \vglue 0.4cm
  Biplob Bhattacherjee$^{(a)}$\footnote{biplob@theory.tifr.res.in} and
  Amitava Datta$^{(b)}$\footnote{adatta@iiserkol.ac.in}
        \vglue 0.1cm
          {\it $^{(a)}$
          Department of Theoretical Physics, Tata Institute of Fundamental Research, \\ 
          1, Homi Bhabha Road, Mumbai 400 005, India. \\
          \it $^{(b)}$
          Indian Institute of Science Education and Research, Kolkata, \\
          Mohanpur Campus, PO: BCKV Campus Main Office,\\
          Mohanpur - 741252, India.
          \\}
          \end{center}
          \vspace{.1cm}
\begin{abstract}
{\noindent \normalsize}

The invariant mass distribution of the di-photons from the decay of 
the lighter scalar Higgs boson(h) to be carefully measured  by 
dedicated h search experiments at the LHC 
may be distorted by the di-photons 
associated with the squark-gluino events with much larger cross sections 
in Gauge Mediated Supersymmetry Breaking (GMSB) models. This distortion 
if observed by the experiments at the Large Hadron Collider at 7 TeV or 
14 TeV, would disfavour not only the standard model but various two 
Higgs doublet models with comparable h - masses and couplings but 
without 
a sector consisting of new heavy particles decaying into photons. The 
minimal GMSB (mGMSB) model constrained by the mass bound on h from LEP and 
that on 
the lightest neutralino from the Tevatron, produce negligible effects. But 
in the currently popular general GMSB(GGMSB) models the tail of the 
above distribution may show  statistically significant excess of events 
even in the early stages of 
the LHC experiments with integrated luminosity insufficient for the 
discovery of h. We illustrate the above points by introducing 
several benchmark 
points in various GMSB models - minimal as well as non-minimal. The 
same
conclusion follows from a detailed parameter scan in a simplified
GGMSB model recently employed by the CMS collaboration to interpret 
their 
searches in the  di-photon + $\etslash$ channel.
Other
observables like the effective mass distribution of the di-photon 
+ X events may also reveal the presence of new heavy particles  beyond 
the Higgs sector. The 
contamination of the h mass peak and simple remedies are also 
discussed.   
\end{abstract}
\begin{center}
%MSSM Higgs, Supersymmetric models,
\end{center}

%-------------------------------------------------------------------------

\newpage

%\maketitle
\section{Introduction}
%\begin{equation}
%{\color{DarkBlue} x = \log_{10} (\nu/\rm MHz) }
%\end{equation}

The search for the Higgs boson/bosons and the study of their 
properties tops the priority list of the on going 
experiments at the LHC  at 7 TeV as well as
of the upcoming experiments  at the highest attainable energy of 14 TeV. 
This will shed light on the mechanism of 
electroweak symmetry breaking (EWSB) and the generation of the particle 
masses. The Higgs sector of the very successful standard model (SM) of 
particle physics responsible for EWSB is indeed very simple. It consists 
of a single neutral, scalar Higgs particle. However, this economical 
Higgs sector turns out to be the Achilles' heel of the SM.

The scale of EWSB is expected to be around 100 GeV. On the other hand 
the naturalness or the hierarchy problem 
\cite{finetuning,gaugehieracrchy} comes into play, should there be some 
new physics beyond the SM characterized by a much higher energy scale. 
The very existence of the new scale tends to push up the EWSB scale far 
beyond the expected magnitude. For example, the Higgs boson mass blows 
up to values closer to the new scale, in contrast to the preferred range 
\cite{EW} $m_h =89^{+35}_{-26}$ suggested by EW precision 
data (the hierarchy problem), unless the parameters of the SM are 
extremely fine tuned (the naturalness problem). Even the absence of any 
compelling evidence in favour of a new physics model does not allow us 
to overlook the above issues, since one cannot wish away the Planck 
scale, the scale of gravity which is always there.

The above theoretical inconsistency has led to several extensions of the 
SM. The minimal supersymmetric standard model(MSSM) \cite{books} removes 
the above problem elegantly in the limit of exact supersymmetry 
(SUSY) and stabilizes the Higgs mass. However, SUSY must be broken by 
the soft breaking terms, since the negative results of sparticle searches at 
various colliders indicate that the superpartners (or the sparticles) 
must be considerably heavier than the corresponding particles. In spite 
of SUSY breaking, the naturalness problem remain under control, provided 
the masses of the sparticles are $\sim$ 1 TeV \cite{books} or smaller.

In the MSSM the Higgs sector must necessarily be extended and the 
spectrum consists of two neutral scalars ($h$ and $H$), one neutral 
pseudo scalar ($A$) and two charged scalars ($H^{\pm}$) 
\cite{books,djouadi}. Enormous amount of work has been done in 
developing the strategies for searching these 
bosons at the LHC \cite{cmstdr,atlastdr}. Adding any new dimension to 
the above studies is not the aim of this paper.    

However,  at the LHC these 
Higgs bosons are likely be produced along with a variety of 
superpartners of the standard model particles - the sparticles. 
In particular the strongly interacting sparticles may indeed be produced
in numbers much larger than that of typical Higgs induced events.
Even if  a small fraction of such events pass through the selection 
criteria for the dedicated search of any 
of the above Higgs 
bosons, then the shape of certain distributions
expected for Higgs production alone may change 
significantly,  indicating that the 
Higgs boson may be part of a framework larger than the SM. Such 
an unexpected shape is likely to 
be prominent near the tails of the distributions under study 
where
contributions from the Higgs signal as well as the SM background are 
naturally small. In this paper we shall illustrate this  possibility 
with a specific example.

Some of the above issues have recently been addressed in the 
context of charged Higgs search by a counting experiment during 
the LHC-14 TeV run \cite{nabanita}. It has been demonstrated 
that both in the unconstrained MSSM and in the minimal 
supergravity(mSUGRA) model \cite{msugra}, SUSY events can 
seriously affect the size of the charged Higgs signal unless 
additional selection criteria are introduced. On the other 
hand, the combined squark-gluino and charged Higgs events may 
establish physics beyond the standard model (BSM) physics at a 
confidence level higher than that attainable by the Higgs 
signal alone. This enhancement would disfavour not only the SM 
but also its 
extensions with two Higgs doublets. After this important hint 
regarding BSM physics, a clean charged Higgs sample may be 
recovered  by additional kinematical cuts. It needs to be 
emphasized that the above issues are relevant in any extension 
of the SM with a new particle sector as well as an extended 
Higgs sector.

In this paper we focus our attention on the search for the 
lighter CP even scalar Higgs boson(h).  The h boson in the MSSM 
has properties very similar to the SM Higgs boson in a large 
volume of the parameter space (the so called decoupling 
region). Thus the discovery of a single neutral Higgs scalar, 
which is a likely scenario in the early stages of the LHC 
experiment, will yield very little information regarding the 
underlying theory. The situation may change dramatically if 
SUSY is realized in nature in such a way that a sizable number 
of sparticle induced events change the characteristics of the expected h 
signal significantly. This can convincingly establish the BSM 
origin of the discovered boson. Even the discovery of one or more 
heavier Higgs boson with no counterpart in the SM, would indicate
only an extended Higgs sector but not the existence of new heavy
sparticles.

The mass of h is always bounded from above ($m_h \leq 140 $ 
GeV) \cite{books} in supersymmetric theories. As a consequence 
the search strategy for it is very similar to that for the 
standard Higgs boson with mass within the above bound 
especially in the decoupling region. If $m_h < 130$ GeV, an 
important discovery channel would be the inclusive di-photon + X 
channel with a relatively modest integrated luminosity ($\lum$) 
of a few tens of $\ifb$  at the 14 TeV run of the 
LHC (see Figure 10.38 of \cite{cmstdr}). Here X stands for any 
other particle accompanying the h signal. This signal stems 
from h production  at the LHC via several channels, followed 
by the one loop decay $h \ra \gamma \gamma$. The suppression of 
the signal due to the tiny branching ratio (BR) of this decay ( 
$\sim 10 ^{-3}$) is adequately compensated by the relatively 
low SM background. In the MSSM also  
a small peak in 
the $\gamma~\gamma$ invariant mass distribution around the h mass would 
establish the h signal (see Figure 11.37 of \cite{cmstdr}).

Among various models of softly broken supersymmetry the Gauge 
Mediated Supersymmetry Breaking (GMSB) models \cite{GMSB,GMSB1} 
(to be briefly reviewed below) predict a large number of 
di-photon + X events at LHC energies coming from the production 
of strongly interacting sparticles. In this paper we wish to 
focus on the impact of these events on the di-photon invariant 
mass distribution to be studied by the dedicated h-search 
experiments with utmost care.

In these models the gravitino with negligible mass turns out to 
be the lightest supersymmetric particle (LSP) and, hence, the 
carrier of missing energy. If the lightest neutralino 
($\lspone$) is the next lightest supersymmetric particle 
(NLSP), it can decay into the LSP and a photon with a large BR. 
In R-parity conserving supersymmetry, sparticles are produced 
in pairs and each of them eventually decay, through a cascade, 
into a gravitino-photon pair. Each SUSY event in this scenario 
will, therefore, have the $2\gamma$ + X topology. Thus the pair 
production of squarks and gluinos in any combination having 
a large cross section may leave their signature in the observed 
$\gamma~\gamma$ invariant mass distribution.

 The Plan of the paper is as follows. In the next section 
we briefly describe different GMSB scenarios and the present 
collider constraints on them. We also define our benchmark points 
for the analysis. In section 3, we present our main results on the
di-photon invariant mass distribution in GMSB models at the LHC.
Our emphasis will
be on the ongoing LHC experiments. However,
the possibilities for the 14 TeV experiments will also be briefly touched
upon. Our conclusions will be summarized in section 4.
 
\section{Different GMSB scenarios, sparticle spectra and the 
current\\ collider constraints}

To begin with we shall review the constraints from different collider 
experiments on the parameter space of the minimal GMSB (mGMSB) model 
\cite{GMSB,GMSB1}. We also wish to add a few points regarding the 
approximations in obtaining these bounds and the resulting uncertainties 
which have not been sufficiently elucidated in the current literature. 
In this model the soft breaking parameters are generated via the gauge 
interactions of the SM. As a result the sparticles with the same SM 
quantum numbers but different flavours acquire the same mass. This keeps 
the potentially dangerous flavour changing neutral current induced 
processes under control.

In the mGMSB model there is an observable sector consisting of the MSSM 
fields. The supersymmetry breaking sector consists of a gauge singlet 
chiral superfield Y, identified with the goldstino superfield in the 
simplest version of the model. The scalar and auxiliary components of Y 
are assumed to develop vacuum expectation values (VEVs) denoted by S and 
F respectively. In the early versions of this model the dynamics of 
generating these VEVs were not specified.  The model also has a 
messenger sector consisting of superfields $\Phi_i$ having the gauge 
interaction of the SM. The messenger superfields interact with Y via a 
tree level interaction in the superpotential. This generates a 
supersymmetric mass of $\cal{O}$(M) of the messenger fields, usually 
referred to as the messenger scale. A SUSY breaking mass squared 
splittings of the order F is also generated within the messenger 
supermultiplets. This SUSY breaking is communicated to the observable 
sector by gauge interactions between the messenger and the observable 
fields via higher order processes.  A restriction in the messenger 
sector comes from the requirement of the unification of the coupling 
constants of the standard model. If the messenger superfields form 
complete GUT multiplets (e.g., 5 or $\bar{5}$ of SU(5)), the value of 
the unification scale or the GUT scale ($M_G$) does not change. However, 
coupling constant unification constrains the number of messenger 
superfields(N) to be $\leq$ 5.

The simple mGMSB model is characterized by five parameters \cite{GMSB,GMSB1} 
\bc
   $\Lambda$, M, $N$, tan $\beta$ and sign($\mu$)
\ec

where $\Lambda$ = F / S is the SUSY breaking scale in the observable 
sector, M is the messenger mass scale, $N$ is the number of messenger 
multiplets belonging to the 5 + $\bar{5}$ representation of SU(5), tan 
$\beta$ is the ratio of the vacuum expectation values of the neutral 
Higgs fields in the observable sector and $\mu$ is the Higgs mixing 
parameter in the superpotential with magnitude fixed by the radiative 
symmetry breaking condition. It bears recall that the soft breaking 
trilinear (A) and bilinear (B) parameters are generated by higher order 
processes and are small at the messenger scale M. The weak scale 
parameters are then obtained by the standard renormalization group (RG) 
evolutions. The sparticle spectrum at the weak scale can be computed by 
both SUSPECT version 2.41 \cite{suspect} and ISAJET \cite{isajet}.
The phenomenology of this model has been discussed by various authors 
\cite{pheno_mgmsb}
 
The current experimental lower bound on the Higgs boson mass from LEP 
$m_h > 114.4$ GeV \cite{lephiggs} poses the strongest constraint on the 
parameter space of the mGMSB model. Given the input parameters one can 
compute the Higgs mass including higher order corrections. In this 
paper $m_h > 114.4$ GeV will be henceforth referred to as the 
stronger bound on the computed h-mass. However, there is an estimated 
theoretical uncertainty of about 3 GeV on the computed h-mass due 
to yet unknown higher order effects\cite{uncertainty}. In view of this 
a point in the parameter space with computed $m_h \geq$ 111.4 GeV 
( the weaker h-mass bound) may still be acceptable. 

%It is well-known that in the MSSM 
%there is an upper bound on $m_h$: $m_h \leq 140.0$ GeV \cite{books}. In 
%mGMSB, however, the mixing parameter in the stop sector turns out to be 
%small because of the small trilinear couplings as discussed above. Thus 
%in this model the above upper bound becomes tighter:$m_h \leq 122.0$ 
%GeV 
%\cite{martin}. This comparatively narrow range of allowed $m_h$ in 
%mGMSB 
%leaves open the possibility of distinguishing this model from mSUGRA 
%via 
%accurate Higgs mass measurements at the LHC even if no direct SUSY 
%signal is observed.

\vskip 10pt
\begin{figure}[htbp]
\centerline{\epsfig{figure=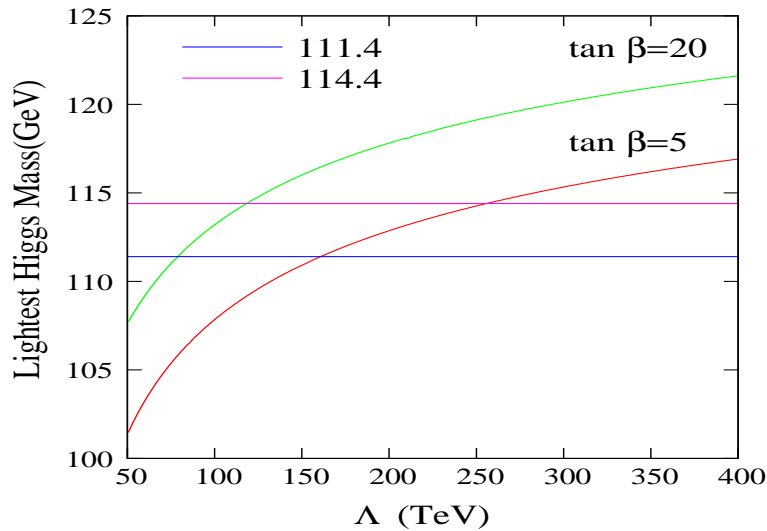,height=7cm,width=10cm,angle=0}}
%\centerline{ \epsfxsize= 6.5 cm\epsfysize=5.0cm  \epsfbox{fig1.eps} }
 \caption{ \footnotesize \it Variation of the computed Higgs mass with 
the SUSY breaking scale $\Lambda$ (in TeV) for different 
tan {$\beta$}=5 (red line) and 20 (blue line) in the mGMSB model.
% in the journal version write bottom to top or solid and dashed lines. 
No significant enhancement of the 
Higgs mass in the region tan$\beta$=20 to 50 has been noticed. The 
horizontal lines correspond to the Higgs mass 111.4 GeV and 114.4 GeV 
respectively. 
The top quark mass is fixed at 173 GeV and M=2 $\Lambda$.}\label{fig:Mh_tb}
\end{figure}

In Figure \ref{fig:Mh_tb} we present $m_h$ computed as a function of $\Lambda$. We 
take $m_t$=173 GeV, $M$=$2 \Lambda$, $N$=1. It follows that for tan 
$\beta$ = 5 (tan $\beta$ = 20) the stronger h-mass bound yields a lower 
bound $\Lambda_{min} \geq$ 257 TeV (118 TeV). If, however, the 
theoretical uncertainties are taken into account and the weaker $m_h$ 
bound is used instead, relaxed bounds $\Lambda_{min} \geq$ 161 TeV (79 
TeV) are obtained. It is interesting to note that for $m_h \approx 120$, 
$\Lambda \approx 300$ GeV which corresponds to gluino mass 
($m_{\tilde{g}}$) and average squark mass ($m_0$) of approximately 2 TeV 
and 3 TeV respectively. Thus should the Higgs mass bound continues to be 
pushed upwards, the squark-gluino search at the LHC, as predicted by 
mGMSB, will be quite challenging.

\vskip 10pt
\begin{figure}[h]
\centerline{\epsfig{figure=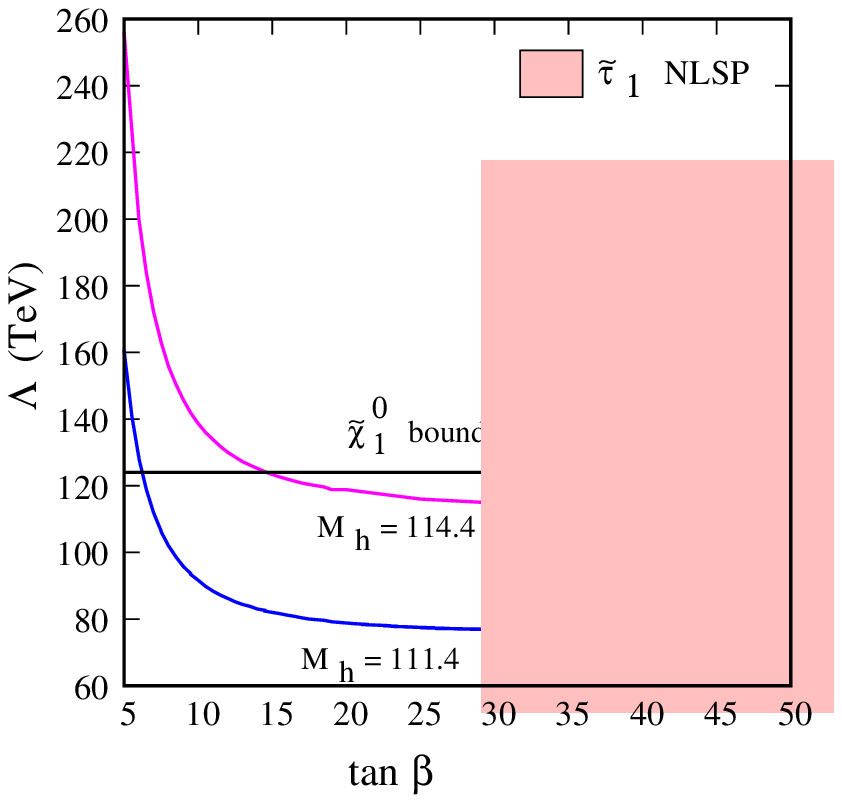,height=7cm,width=10cm,angle=0}}
%\centerline{ \epsfxsize= 6.5 cm\epsfysize=5.0cm  \epsfbox{fig1.eps} }
 \caption{ \footnotesize \it Allowed parameter space in the mGMSB model in 
the $ \Lambda$ (in TeV)- tan{$\beta$} plane.
 The other parameters are as in Figure 1. In the shaded (pink) region, 
$\tilde{\tau}_1$ is the NLSP. The magenta and blue 
lines correspond to the Higgs mass 114.4 GeV and 111.4 GeV respectively. 
The horizontal black line indicates the 
latest neutralino mass bound set by $D\O{}$ collaboration. }\label{fig:scan}
\end{figure}

The $D\O{}$ and CDF collaborations have put strong constraints on NLSP 
($\chi_1^0$) mass. In order to obtain quantitative results they consider 
SPS8 slope in the mGMSB model which has only one parameter $\Lambda$. 
The other parameters are fixed and given by $M$=2$\Lambda$, $\tan 
\beta$=15, $N$=1 and $\mu>0$. The lifetime of the $\chi_1^0$ is not 
fixed by these parameters and it is assumed to be sufficiently short. It 
means that the photon coming from the decay of $\chi_1^0$ is prompt. The 
gaugino pair production processes are expected to be dominant at the 
Tevatron. The decay of two $\chi_1^0$ produce two photons and gravitinos 
that give rise to missing energy.

The latest constraint is from $D\O{}$ collaboration who have obtained 
a lower bound  124 TeV on the scale $\Lambda$ with 
6.3 
$\ifb$ of data at Run II \cite{D0}. We have computed the leading order 
(LO) cross sections for the dominant SUSY processes at Tevatron energies 
(mainly 
electroweak gaugino pair production) using PYTHIA \cite{pythia}. The
next to leading order(NLO) cross section and 
the K factor with renormalization scale($\mu_R$)= factorization scale 
($\mu_F$)= 2$\mchonepm$ is computed by PROSPINO \cite{prospino, NLO}.
Our cross section for 
$\Lambda$=124 TeV (assuming SPS8 point) is in good agreement with the 
result in \cite{D0}. We find that the above bound corresponds to an
upper bound on the 
SUSY production cross section of 4.5 fb. For promptly decaying 
$\lspone$ this $\Lambda_{min}$ yields $m_{\lspone} \geq$ 174 GeV. This 
bound is indicated in the $\Lambda$ - tan$\beta$ plane 
(Figure \ref{fig:scan}) by the 
horizontal line. For comparison the stronger and weaker h-mass bounds 
are also indicated in the same figure. It follows that the D0 bound
on $\Lambda_{min}$ supersedes that from the stronger (relaxed) $m_h$ 
bound for tab $\beta \gsim$ 15 (6.5). It follows that irrespective of 
the uncertainties in $m_h$ due to higher order corrections and the 
choice of tan $\beta$, values of $\Lambda$ 
smaller than 124 TeV are disfavoured. This bound also supersedes the 
earlier CDF lower bound on SUSY scale 107 TeV\cite{CDF}. The assumptions 
underlying this bound will be critically examined below.  

Strictly speaking the $D\O{}$ bound is valid for the snowmass slope 
SPS8 with variable $\Lambda$. The other parameters are M = 2 
$\Lambda$, $N$ = 1, tan 
$\beta$ = 15 and sign ($\mu) >$ 0. For $N \geqslant$ 2, the scenario reduces
to a $\tilde{\tau}$ NLSP scenario which is not under consideration
in the present paper.  

However, keeping other parameters
fixed and varying tan $\beta$ we have checked that neither the cross 
sections
of the dominant processes( $\chplus_1 - \lsptwo$ and $\chplus_1 - 
\chminus_1$
pair production \cite{D0,CDF}) at the Tevatron energy  nor the $\chplus$ 
and 
$\lsptwo$ masses vary appreciably with tan 
$\beta$. This is, however, expected since over the scanned  parameter 
space the above gauginos are wino like to a very good approximation and 
have masses controlled by the $SU(2)$ Gaugino mass $M_2$ at the weak 
scale. The cross section, therefore, shows only a mild 
tan $\beta$ dependence. 

The bino-like $\lspone$ mass on the other hand is determined by
the $U(1)$ gaugino mass parameter ($M_1$) at the weak scale related to  
$M_2$ by the gaugino mass unification condition. We 
shall, therefore, use the $D\O{}$ bound throughout this analysis unless the 
electroweak gauginos happen to be  significantly mixed, which is often the 
case for non-minimal models to be discussed below.

\begin{figure}[htbp]
\centerline{\epsfig{figure=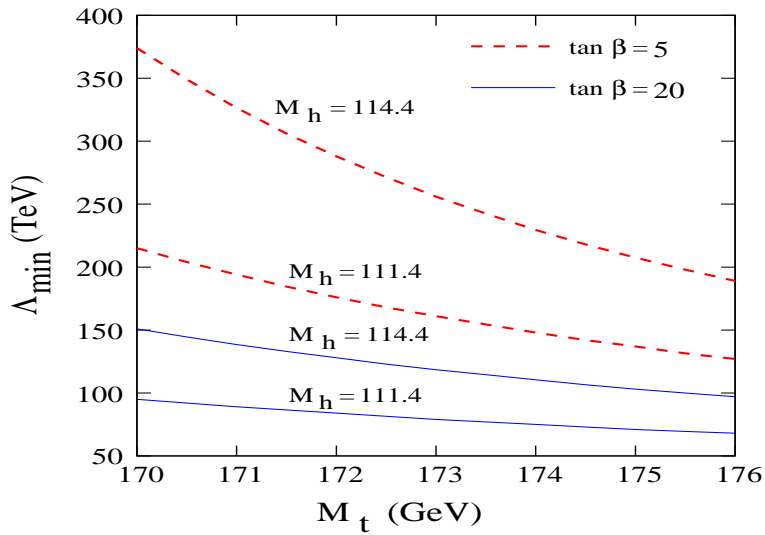,height=7cm,width=10cm,angle=0}}
%\centerline{ \epsfxsize= 6.5 cm\epsfysize=5.0cm  \epsfbox{fig1.eps} }
 \caption{ \footnotesize \it Minimum allowed value of the SUSY scale 
$\Lambda$ as a function of the top mass obtained from the stronger and
weaker Higgs mass bound for two values of tan$\beta$.}
 \label{fig:top_sen}
\end{figure}

The neutralino mass bound from Tevatron has an important advantage. 
The predicted $m_h$
depends sensitively on $m_t$ (recall the $m_t^{4}$ dependence in the 
radiative corrections to $m_h$). The minimum value of  
$\Lambda$ consistent with the weaker or stronger h-mass bound  
as a function of $m_t$ is presented in Figure \ref{fig:top_sen} for two values of tan
$\beta$: 5 and 20. The top masses on the X-axis are allowed by the
current bound obtained by a recent CDF measurement 
$m_t = 173.13 \pm 1.2$ \cite{topmass}.

These curves illustrate the uncertainty in 
$\Lambda_{min}$ due to the present error in  $m_t$.
The neutralino 
mass bound, on the other hand, shows no strong 
$m_t$-dependence. Throughout this paper we shall use $m_t$ = 173 GeV. 

In summary,
while choosing the representative points for 
calculating the SUSY contribution to the di-photon invariant mass 
spectrum  in the 
GMSB model, the $\Lambda_{min}$ from the weaker h-mass bound (neutralino 
mass bound ) should be used for tan $\lsim$ 6.5 ( tan $>$ 6.5). 

Another theoretical uncertainty comes from the choice of the messenger 
scale M. In Table \ref{m_variation} we present the variation of the sparticle spectrum 
with M. We find that the variation of M within moderate ranges does not 
affect the spectrum and, hence, the collider signatures drastically. 
\vskip 10pt 
\begin{table}[t]\footnotesize
\begin{center}
\scalebox {1 }[0.90]{
\begin{tabular}{|c|c|c|c|c|c|c|}
\hline
Masses &CASE I          &CASE II         &CASE III      \\  
       &tan$\beta$=20    &tan$\beta$=20    &tan$\beta$=20   \\ 
       &$\Lambda$=120   &$\Lambda$=120  &$\Lambda$=120 \\
       &$M_m$=2$\Lambda$&$M_m$=5$\Lambda$&$M_m$=10$\Lambda$ \\
\hline
$\mgl$           &    997.6    &    963.8    &     959.1    \\
$\mul$           &   1320.4    &   1293.6    &    1280.3    \\
$\mur$           &   1264.4    &   1235.8    &    1221.2     \\
%$\mdl$           &   1322.7    &   1295.9    &    1282.7     \\
% $\mdr$          &   1259.4    &   1230.6    &    1215.8     \\
% $\mbone$        &   1246.5    &   1215.7    &    1199.6     \\
% $\mbtwo$        &   1275.2    &   1243.5    &    1226.9     \\
% $\mtone$        &   1155.4    &   1115.3    &    1092.6     \\
% $\mttwo$        &   1281.1    &   1249.2    &    1232.6     \\
$\mselecnu$      &    417.4    &    420.3    &     423.6    \\
$\mselecl$       &    424.6    &    427.5    &     430.8    \\
 $\mselecr$      &    209.7    &    213.3    &     216.4  \\
% $\mstaunu$      &    416.2    &    418.9    &     422.1  \\
 $\mstauone$     &    199.9    &    202.2    &     204.6  \\
% $\mstautwo$     &    425.9    &    428.9    &     432.1  \\
$\mlspone$       &    169.1    &    162.8    &     161.9   \\
$\mlsptwo$       &    321.4    &    311.8    &     311.1   \\
%$\mlspthree$     &   -493.0    &   -522.2    &    -539.9    \\       
%$\mlspfour$      &    511.5    &    536.8    &     553.1   \\
$\mchonepm$      &    321.1    &    311.7    &     310.9   \\
%$\mchtwopm$      &    511.6    &    537.1    &     553.5    \\
 $\mhiggsone$    &    114.5    &    114.3    &     114.1    \\
%$\mhiggstwo$     &    610.8    &    634.2    &     650.1   \\
%$\mhiggsthree$   &    610.6    &    633.9    &     649.9    \\
% $\mchhiggs$     &    616.3    &    639.4    &     655.3    \\
\hline
\end{tabular}
}
\vskip 5pt
\caption{\footnotesize \it Variation of sparticle masses as a function of 
the Messenger scale ($M_m$) for 
fixed SUSY scale and $\tan \beta$ in the minimal GMSB model.}\label{m_variation}
\end{center}
\end{table}

It should be emphasized that the CDF and $D\O{}$ experiments actually 
constrain the quantity $\sigma_{SUSY} \times BR (\lspone \rightarrow 
\gamma~~\tilde{G}) ^2$, where $\sigma_{SUSY}$ is the total sparticle 
production cross section within the kinematic reach of the Tevatron, BR 
($\lspone \rightarrow \gamma~~\tilde{G})$ is the branching ratio 
underlying each single photon event. The cross section limit quoted 
above is, therefore, subject to the model dependent assumption BR 
($\lspone \rightarrow \gamma~~\tilde{G})^2 ~\approx 1$.

The above BR depends on the neutralino mixing angles and may differ 
significantly from 1, if the lightest neutralino is indeed an admixture 
of 
electroweak gauginos and higgsinos. The widths of the different decay 
channels of $\lspone$ are given, e.g., in Eqs 3.7 - 3.12 of 
\cite{GMSB1}. The relative width of the $\lspone$ decay into $\gamma$ 
and Z is given by (following the notation of \cite{GMSB1})

\bc
$\Gamma (\lspone \rightarrow Z ~~\tilde{G}) = (\kappa_Z/~ 
\kappa_{\gamma})^2~\Gamma 
(\lspone \rightarrow \gamma~~\tilde{G}) \times (1 - (m_Z/ 
m_{\lspone})^2)^2$\\
\ec
\vskip 10pt
\begin{table}[t]
\begin{center}
\begin{tabular}{|c|c|c|c|}
\hline
$\Lambda$(TeV) & BR ($\lspone \rightarrow \gamma~~\tilde{G})$ \\
\hline
 105     &       0.950 \\
 125     &       0.907 \\
 145     &       0.874 \\
 165     &       0.849  \\
\hline
\end{tabular}
\caption{\footnotesize \it Branching fraction of $\lspone \rightarrow
\gamma \tilde{G}$ for $\tan \beta$=20 
and $M$= 2$\Lambda$ in the mGMSB model. }\label{branching}
\end{center}
\end{table}

As long as the $\lspone$ is a pure Bino, which is almost always the case 
in mGMSB, the ratio ($\kappa_Z/~ \kappa_{\gamma})^2 \approx tan^2 
\theta_W $. Moreover for relatively small $\Lambda$, $m_{\lspone} 
\approx m_Z$, and the Z-channel is strongly suppressed compared to the 
$\gamma$ channel. However, as $\Lambda$ and $m_{\lspone}$ increases, the 
relative importance of the Z-channel increases even in the mGMSB 
model, 
due 
to 
the factor $(1 - (m_Z/ m_{\lspone})^2)^2$. As a result the BR ($\lspone 
\rightarrow \gamma~~\tilde{G})$ becomes significantly smaller than one. 
This is illustrated in Table \ref{branching}. For $\Lambda \approx 
125$ TeV, which is the current limit from the Tevatron, BR ($\lspone 
\rightarrow \gamma~~\tilde{G}) \approx 0.91$. This will relax the limit 
on $m_{\lspone}$ by a few GeV only. However, as the Tevatron experiments 
become sensitive to higher $\Lambda$, the value of BR ($\lspone 
\rightarrow \gamma~~\tilde{G}$) should be carefully folded into the 
calculation before extracting the $\mlspone$ limit. In the following 
we shall also consider more general GMSB models where the value of the 
BR is considerably smaller than unity and must be taken into account 
before extracting any limit from the Tevatron data.

\begin{table}[htbp]
\begin{center}
\begin{tabular}{|c|c|c|c|}
\hline
          {  BP I   }                               &     {   BP II}            \\
\hline
$\Lambda$=165 TeV, $M_m$= 2$\Lambda$                   &    $\Lambda$=125 TeV $M_m$= 2$\Lambda$  \\      
$\tan \beta$=5, $N_1=N_2=N_3=1$                        &     $\tan \beta$=20, $N_1=N_2=N_3=1$    \\
$m_h=$111.6 GeV                                        &     $m_h=$114.8 GeV                      \\
\hline
\end{tabular}
\caption{\footnotesize \it Benchmark points in the minimal GMSB model.}\label{BP_mGMSB}
\end{center}
\end{table}

In view of the above discussions we have chosen the points in Table \ref{BP_mGMSB} 
which are consistent with the constraints discussed above, 
for estimating the  squark gluino contribution to the $\gamma - \gamma$
invariant mass distribution from di-photon + X events in the mGMSB model.

For point I with tan $\beta =$ 5, the computed $m_h$ is consistent with 
the weaker bound and $\Lambda = 
\Lambda _{min}$ as obtained from  Figure \ref{fig:scan}.  
This point yields gluino mass ($m_{\tilde{g}}$) = 1329 GeV and 
average squark mass ($m_{\tilde{q}}$) = 1725 GeV. For Point II  
$\Lambda = \Lambda_{min}$ as obtained from the $m_{\lspone}$ bound, 
since tan $\beta = 20$. This leads to $m_{\tilde{g}}$ = 
1035 GeV and $m_{\tilde{q}}$ = 1334 GeV. The squark-gluino contributions 
to the di-photon + X events in these scenarios will be computed in the 
next section.

In the mGMSB model a single scale ($\Lambda$) controls both the sfermion 
and 
gaugino masses. As a result these masses are of the same 
order. Over the years several extensions of the mGMSB models have been 
proposed. We shall collectively call them  non-minimal models. 
For phenomenological studies such extensions can be realized 
by introducing new  parameters in addition to the set 
of five parameters mentioned above.  

One possibility is that the gaugino 
masses may be suppressed
with respect to the sfermion masses \cite{lowscale}. This scenario can 
be studied by introducing a  parameter R ($<1$) such 
that $\Lambda_G = R \Lambda_S$, where $\Lambda_G$($\Lambda_S$) is the 
scale in the gaugino ( scalar ) sector at the messenger scale.     

More recently theoretically well-motivated extensions of the mGMSB  
model reflecting the above feature have been constructed.
In a wide class of models the gaugino mass scale ($\Lambda_G$)
happens to be  severely suppressed in comparison to the sfermion mass 
scale ($\Lambda_S$)\cite{small}. This suppression is a consequence of
expanding around the lowest classical vacuum of the low energy effective 
theory \cite{metastable}.  
This can be avoided, e.g., if the vacuum is an excited metastable 
state. It is, therefore, possible to construct  gauge mediation models 
where the ratio of gaugino and sfermion masses continuously vary from 
very small to large values \cite{abel}.
 
The sparticle spectra and BRs for different R can 
be computed by using, for example, online version of ISAJET. For $R < 1$ the gluino mass 
is expected to be significantly  smaller 
than that in the mGMSB model (R =1) with $\Lambda$ = $\Lambda_S$. 
However, the chargino and the neutralino masses will also be 
correspondingly suppressed in the above model. Hence, the Tevatron bound does not allow arbitrarily 
small $R$ as we shall see below.

For example, with $\Lambda_S$=124, which corresponds to the Tevatron lower bound in the mGMSB model (see 
Point II) , we obtain for R = 0.6, $\mlspone = 101$, $\mlsptwo = 205 \approx m_{\chone^{\pm}}$, $\mgl$=667 and 
BR ($\lspone \rightarrow \gamma~~\tilde{G}$) =0.98. It is obvious that with such light electroweak gauginos, 
the combind electroweak gaugino production cross section at Tevatron is 
enhanced violating the 
bound. For tan $\beta$ = 5 (20) the modified bound turns out to be $\Lambda_S > 203 (202)$ TeV. The Higgs masses 
in this two case are 112.7(117.5) GeV which is above the weaker $m_h$ bound. Thus in contrast to the mGMSB model, 
the lower bound on $\Lambda$ comes from the Tevatron data for both low and high tan $\beta$. It is now easy to 
check the lower bounds on the gluino and average squark masses. They are $\mgl \approx$ 1030 GeV   
and $m_{\tilde{q}} \approx$ 2 TeV. Thus  although the minimum gluino mass is less than or comparable to the 
corresponding values in the mGMSB model, the average squark mass is much larger. Thus in general the size 
of the total squark-gluino signal and, consequently, the SUSY induced di-photon signal is expected 
to be smaller in non-minimal models with $R < 1$  
compared to the corresponding mGMSB model.

It has further been pointed out that not only the overall mass scales
in the gaugino and sfermion sectors may be  different, the  $SU(3), 
SU(2)$ 
and $U(1)$ contributions of messenger fields  to the masses of 
the gauginos and the 
sfermions having different gauge quantum numbers, may also differ 
from each other. This happens, e.g,  when the
messenger fields do not belong to a complete multiplet of a GUT group
(say, $SU(5)$) \cite{lowscale}.
\begin{table}[t]\footnotesize
\begin{center}
\scalebox {1 }[0.85]{
\begin{tabular}{|c|c|c|c|c|c|c|}
\hline
      & { BP III}   & { BP IV }   &{ BP V}      & { BP VI}   & { BP VII}  & { BP VIII}  \\
\hline
       &tan$\beta$=5    &tan$\beta$=20   &tan$\beta$=5    &tan$\beta$=20  &tan$\beta$=5   &tan$\beta$=20   \\
       &$R_{sl}$=0.4    &$R_{sl}$=0.4    &$R_{sl}$=0.6    &$R_{sl}$=0.6   & $R_{sl}$=1    &$R_{sl}$=1      \\
Masses &$\Lambda$=160   &$\Lambda$=170   &$\Lambda$= 145  &$\Lambda$=138  &$\Lambda$=132  &$\Lambda$=96    \\
       &$N_1$=$N_2$=2,  &$N_1$=$N_2$=2,  &$N_1$=$N_2$=2,  &$N_1$=$N_2$=2, &$N_1$=$N_2$=2, &$N_1$=$N_2$=2,  \\
       &$N_3=1$         &$N_3=1$         &$N_3=1$         &$N_3=1$        &$N_3=1$        &$N_3=1$         \\
\hline
$\mgl$          &  601.43  &  634.84   &  769.25   &  735.97   & 1080.68   &  811.35    \\ 
$\mul$          & 1707.57  & 1807.49   & 1576.26   & 1505.46   & 1512.87   & 1125.34   \\
$\mur$          & 1548.76  & 1637.86   & 1432.95   & 1368.84   & 1382.82   & 1032.09   \\
$\mdl$          & 1709.33  & 1809.28   & 1578.16   & 1507.62   & 1514.84   & 1128.21    \\ 
$\mdr$          & 1532.50  & 1620.46   & 1418.62   & 1355.44   & 1370.62   & 1023.95    \\ 
$\mbone$        & 1543.42  & 1617.90   & 1427.69   & 1352.12   & 1377.27   & 1019.73    \\ 
$\mbtwo$        & 1636.31  & 1728.66   & 1511.61   & 1440.96   & 1454.46   & 1080.55    \\  
$\mtone$        & 1380.99  & 1468.53   & 1280.93   & 1231.20   & 1245.24   &  937.82    \\ 
$\mttwo$        & 1649.96  & 1742.71   & 1525.31   & 1454.71   & 1468.71   & 1095.36    \\ 
$\mselecnu$     &  779.80  &  829.17   &  715.96   &  682.56   &  679.52   &  494.71    \\ 
$\mselecl$      &  789.04  &  838.58   &  725.12   &  691.94   &  688.63   &  504.75   \\  
$\mselecr$      &  389.25  &  414.11   &  355.88   &  339.15   &  332.85   &  243.13    \\ 
$\mstaunu$      &  779.64  &  826.54   &  715.82   &  680.40   &  679.39   &  493.20   \\   
$\mstauone$     &  388.55  &  399.75   &  355.00   &  327.04   &  331.61   &  234.25    \\  
$\mstautwo$     &  788.82  &  835.52   &  724.92   &  689.48   &  688.44   &  503.11    \\ 
$\mlspone$      & -174.37  & -185.26   & -236.53   & -221.43   & -332.55   & -216.57    \\ 
$\mlsptwo$      & -332.91  & -325.56   & -377.98   & -309.82   &  372.92   &  249.46    \\  
$\mlspthree$    &  441.64  &  375.01   &  406.32   &  323.65   & -398.81   & -286.55   \\  
$\mlspfour$     & -477.58  & -441.27   & -524.44   & -490.80   & -744.13   & -549.85    \\
$\mchonepm$     & -325.08  & -320.00   & -369.82   & -304.86   & -366.79   & -240.69   \\
$\mchtwopm$     & -481.08  & -443.14   & -524.45   & -484.62   & -735.66   & -543.03   \\
$\mhiggsone$    &  111.38  &  116.65   &  111.01   &  115.70   &  111.02   &  114.06   \\
$\mhiggstwo$    &  908.35  &  860.75   &  832.37   &  711.81   &  778.70   &  516.90   \\
$\mhiggsthree$  &  907.39  &  860.54   &  831.35   &  711.61   &  777.63   &  516.72    \\
$\mchhiggs$     &  911.02  &  864.64   &  835.29   &  716.49   &  781.82   &  523.31   \\
\hline
\end{tabular}
}
\vskip 35pt
\caption{\footnotesize \it Spectra for benchmark points in different 
non-minimal GMSB 
scenarios.}\label{BP_nonminimal}
\end{center}
\end{table}
\vskip 10pt

Let us concentrate on a phenomenological model characterized by 5 
parameters in addition to M, tan $\beta$ and sign($\mu$): R (defined 
above), $\Lambda_S$, $N_1$, $N_2$ and $N_3$. In the limit R = 1 and 
$N_1$= $N_2$= $N_3$ = $N$, this model reduces to mGMSB. The explicit 
mass formulae for the sfermions and the gauginos and further references 
can be found in Appendix A of \cite{lowscale}.

The sparticle spectrum in this model can be computed by ISAJET. The 
Bino, Wino and gluino masses at the messenger scale are proportional to 
$N_1, N_2$ and $N_3$ respectively. Clearly if $N_1, N_2 > N_3$, the mass 
difference between the electroweak gauginos and gluinos at the 
electroweak scale obtained by renormalization group evolution, will be 
reduced. Thus in this scenario the Tevatron bound becomes compatible 
with relatively light gluinos and, hence, one obtains a larger number of 
SUSY induced di-photon events. We illustrate some representative 
sparticle spectra with several benchmark points
 ({ BP III} - { BP VIII}) in Table \ref{BP_nonminimal}.

For each point in Table \ref{BP_nonminimal} the dominant chargino-neutralino pair 
production cross section ( NLO ) as a function of $\Lambda$  
is computed with PROSPINO assuming $\mchonepm \approx \mlsptwo$. The 
K-factors for some representative values of $\mchonepm$ are given
in Table \ref{k_factor}. Comparing the computed value with the
the cross section upper bound from Tevatron data, we obtain 
$\Lambda_{min}$ in each case.
Each  $\Lambda_S$ presented in Table \ref{BP_nonminimal} is above the 
corresponding 
$\Lambda_{min}$.  

\vskip 10pt
\begin{table}[h]
\begin{center}
\begin{tabular}{|c|c|c|c|}
\hline
 $\mchonepm$ & K- factor \\
\hline
 100     &      1.39  \\
 200     &      1.29  \\
 300     &      1.15  \\
 400     &      0.99   \\
\hline
\end{tabular}
\caption{\footnotesize \it The K factors of $\chi_1^{+} \chi_2^0$ 
production at the Tevatron for different $\mchonepm$. 
We choose mass of the final state as the QCD scale. }\label{k_factor}
\end{center}
\end{table}

It may be noted from Table \ref{BP_nonminimal} that for most of the non-minimal models of 
this type, the gluino masses are significantly smaller than one TeV. 
Another point of phenomenological interest is the composition of the 
electroweak gauginos. In the non-minimal model $\mu$ as determined by 
the electroweak symmetry breaking condition often turns out to be 
comparable or even smaller than $M_1$. As a result the higgsino 
component of the NLSP changes significantly. In Table \ref{cross} we present $M_1, 
M_2$ and $\mu$ in each model. The BR ($\lspone \rightarrow 
\gamma~~\tilde{G}$) is also shown. It follows that whenever the NLSP 
develops a significant higgsino component, channels like $\lspone 
\rightarrow Z ~~\tilde{G}$, $\lspone \rightarrow h~~\tilde{G}$ open up 
and BR ($\lspone \rightarrow \gamma~~\tilde{G}$) decreases. This has 
been carefully taken into account in extracting limits from the Tevatron 
data as well as in computing the size of the di-photon signal at the LHC 
in different scenarios.

General gauge mediated symmetry breaking (GGMSB) models 
have been proposed recently \cite{ggmsb}. These theoretically 
well-motivated models incorporate the  features of the above 
phenomenological models. A large number of authors have 
recently studied the phenomenology of GGMSB \cite{phenoggmsb}. In such a 
general framework, all gaugino mass parameters are independent of each 
other and squark masses and gluino mass are not correlated. Thus squark 
masses can be lighter in contrast to the models discussed above. To 
illustrate this point, we take two benchmark points where squark masses 
are free parameters. In { BP IX}, we take the point { BP III} and 
set all squark masses to be 1 TeV. The { BP X} point is identical to 
{ BP V} with $m_{\tilde{q}}$ = $m_{\tilde{g}}$.

\vskip 10pt
\begin{table}[htbp]\footnotesize
\begin{center}
\begin{tabular}{|c|c|c|c|c|c|c|c|c|}
\hline
Points             &$M_{\tilde q}$ &$ M_{\tilde g}$ & $M_1$ & $M_2$  & $\mu$   & Total NLO cross sec &  Br($\chi_1 \rightarrow \gamma + \tilde{G} $) & Effective cr-sec \\
%                  &      &        &       &        &         &                     &                  \\
\hline   %         &      &        &       &        &         &             &                           \\
{ BP I}       & 1725 &1329    &  240  &  445   &    884  &   221  fb ( 19 fb)  &   0.85  &  160 fb  ( 14 fb) \\
~{ BP II}     & 1334 &1035    &  181  &  339   &    502  &   1.1  pb ( 84 fb)  &   0.91  &  910 fb  ( 70 fb) \\
~~{ BP III}   & 1618 &601     &  181  &  349   &    432  &   10.5 pb (583 fb)  &   0.89  &   8.3 pb (462 fb) \\
~{ BP IV}     & 1509 &635     &  192  &  371   &    363  &   8.7  pb (414 fb)  &   0.87  &   6.6 pb (313 fb) \\
{ BP V}       & 1495 &769     &  247  &  474   &    397  &   3.0  pb (112 fb)  &   0.81  &   2.0 pb ( 74 fb) \\    
~{ BP VI}     & 1427 &736     &  235  &  452   &    313  &   4.1  pb (207 fb)  &   0.79  &   2.6 pb (129 fb) \\    
~~{ BP VII}   & 1439 &1081    &  376  &  718   &    364  &   725  fb ( 48 fb)  &   0.38  &   105 fb  (7 fb)  \\    
~~~{ BP VIII} & 1072 &811     &  273  &  525   &    240  &   4.8  pb (319 fb)  &   0.35  &   590 fb  (39 fb) \\    
~{ BP IX}     & 1000 &601     &  181  &  349   &    432  &   16.8 pb (822 fb)  &   0.89  &  13.3 pb  (651 fb) \\    
{ BP X}       & 770  &770     &  247  &  474   &    397  &   11.0 pb (640 fb)  &   0.81  &   7.2 pb  (420 fb) \\   
\hline
\end{tabular}
\caption{\footnotesize \it The average squark mass, gluino mass, total NLO SUSY production cross 
section at the LHC for $\sqrt{s}$=14 TeV (7 TeV) and  BR($\chi_1 \rightarrow \gamma + \tilde{G} $) for benchmark points ({ BP I} to { BP X}). 
The elctroewak gaugino mass parameters $M_1$, $M_2$ and $\mu$ are also given 
. Note that in case of 
{ BP I}, { BP II} and { BP VII}, the EW gaugino production cross sections are are comparable to 
the strong production cross sections. }\label{cross}
\end{center}
\end{table}

The total SUSY cross section at the LHC in each scenario is also 
presented in Table \ref{cross}. The leading order and next to leading order (NLO) 
squark gluino cross sections are computed by PROSPINO \cite{prospino}. 
The relatively modest electroweak production cross sections are computed 
by PYTHIA and 
multiplied by the appropriate K 
factor, as discussed above. In each case the QCD scale is chosen 
to be equal to the mass or the average mass  
of the sparticles in the final state and CTEQ 5L and CTEQ 5M parton 
density functions have been used for LO and NLO cross sections 
respectively. 
The cross sections agree well with ref 
\cite{Beenakker:2009ha}.

It is found that the scenarios with relatively light gluinos ({ BP 
III - VI}) have significantly larger cross sections compared to the 
representative mGMSB scenarios ({ BP I} and { BP II}). However, 
for R = 1, ( { BP VII} and { BP VIII}) the NLSP has substantial 
higgsino component. As a result  the relevant BRs are also small in 
the corresponding models and, consequently, the number of the SUSY 
di-photon events 
is rather small as we shall see in the next section.

Recently the CMS collaboration has reported negative results for
squark-gluino search \cite{cmsdiphoton}
search in the di-photon + $\etslash$ channel in the context of
a simplified GGMSB model \cite{lhcwg}. In this model the
$m_{\tilde{\chi_1^0} } $ (the NLSP mass),
$M_{\tilde{q}}$ and $M_{\tilde{g}}$ at the weak scale are taken as 
variables whereas all
all other sparticle masses are fixed at 1.5 TeV.
The impact of the squark-gluino induced di-photons in this case 
on the $\gamma~\gamma$ invariant mass data collected by dedicated Higgs 
search experiments in the context of this model will also be reported in 
the next section.

\section{ Footprints of squark-gluino 
events in the di-photon invariant mass data collected by the Higgs 
search experiments at the LHC.}

From the sparticle spectra presented in the last section we find that in 
all cases the pseudoscalar Higgs boson mass ($M_A$) is much larger than 
$M_Z$. Hence we are in the deep decoupling regime. As a result the 
production cross sections and the BRs of the h-boson are practically the 
same as the SM Higgs boson having the same mass. For h-bosons with 
masses a few GeV above the LEP lower bound, the di-photon+ X channel is 
the most promising signal.

As already mentioned in
the introduction  the prospect of discovering the h-signal in
the di-photon+ X channel at the LHC, is not the main  concern of this
paper. This has been dealt with in great
depth by the LHC collaborations \cite{cmstdr,atlastdr}. Our main task is
to study the
possibility of distortion in the tail of the $\gamma - \gamma$ 
invariant mass
distributions in different GMSB scenarios in typical Higgs search
experiments. For this we simulate by PYTHIA the SM
backgrounds and squark-gluino events
in the models discussed in Section 2 using
the cuts employed by Higgs search experiments.

We begin by following the standard selection procedures
for Higgs search in the di-photon + X
channel by the CMS collaboration \cite{cmstdr}
at 14 TeV.
We note that ATLAS collaboration \cite{atl_conf}
has used same cuts both for 7 and 14 TeV
analyses. Thus our approach  seems to be reasonable.

We require events with  exactly  two isolated photons with 
pseudo-rapidity 
$|\eta| < $ 2.5 and $p_T$ greater than 40 GeV and 35 GeV respectively.
We put the following isolation conditions on photons.
\begin{enumerate}
\item No charged particles with $p_T$ larger than 1.5 GeV/c should be present inside a cone
with $\Delta R <$ 0.3 around the photon candidate where $\Delta R$ = $ \sqrt {\Delta \eta^2 + \Delta \phi^2}$.
\item The total $E_T$ of electrons or photons with 0.06 $< \Delta R <$ 0.35 around the direction
of the selected photon candidate, must be less than 6 GeV in the barrel and 3 GeV in the endcaps.
%(1-1.479)and 3 GeV in the endcaps(1.479 to 3) 
\item The total transverse energies of hadrons within $\Delta R <$ 0.3 around the photon
candidate  must be less than 6 GeV in the barrel and 5 GeV in the endcaps.\\
%(0-1.4) (1.4 to 3)
%\item $P_T$ cut  \cite{atl_conf}.
%\item other cuts \cite{atl_conf}.
%\item The isolation of the photon tracks is often based on detector 
%simulations. For our simple minded generator level analysis we
%follow the prescription of the cms collaboration \cite{cmshiggs}. We 
%require 
\end{enumerate}

The irreducible SM backgrounds  are from   
i)$q$ $\bar{q}$ $\rightarrow$ $\gamma$ $\gamma$ and 
ii) $g$  $g$ $\rightarrow$ $\gamma \gamma$ events. We generate these 
backgrounds by PYTHIA 6.4.21 subject to the above cuts.
The LO cross sections of the backgrounds  are computed by PYTHIA.  
The K factors of these two processes are given in \cite{cmstdr}  
for 14 TeV LHC(1.5 and 1.2 respectively). 
The NLO cross sections for the SM backgrounds at 7 TeV are not 
available in the literature. As a reasonable guess we take the same 
K factor for i) and (ii) to be 1.5 and 1.2. 

The additional cuts related to 
photon reconstruction employed by the CMS collaboration  cannot be 
implemented in our analysis with the toy 
detector of PYTHIA. We assume that their absence will affect the 
SUSY events 
and the background similarly and our main conclusions will be by and 
large valid. We have also ignored the instrumental backgrounds. 
We admit that the main purpose of this analysis is to illustrate 
the possibility of 
distortion in the shape of an expected  distribution due to new 
physics and not to present very accurate numerical results.

The sparticle spectra in different non-minimal models (BP III - VIII in 
Table \ref{BP_nonminimal}) can be readily generated by online ISAJET \cite{isajet}.
The spectra in mGMSB (BP I and II in Table \ref{BP_mGMSB}) and GGMSB models (BP IX 
and X ) have been generated by SUSYHIT \cite{SUSYHIT}. Finally the 
spectra are interfaced with PYTHIA for event generation. For 
squark-gluino events the NLO cross sections are directly computed by 
PROSPINO (see section 2).
\begin{table}[h]\footnotesize
\begin{center}
\begin{tabular}{|c|c|c|c|c|c|}
\hline
Processes& $M_{\gamma \gamma} > 200$ & $M_{\gamma \gamma} > 300$ & $M_{\gamma \gamma} > 400$& $M_{\gamma \gamma} > 500$ & $M_{\gamma \gamma} > 600$ \\
\hline
Born (SM bg)     & 707.6     & 219.5      & 91.3      & 42.8      &20.6  \\
\hline
Box  (SM bg)     &  78.9      &12.8     &  2.5       &0.6       &0.1  \\
\hline
{ BP I}       &   5.5      & 3.0     &  1.5       &0.7    &   0.3  \\
\hline
{ BP II}      &  18.2      & 7.7     &  3.1       &1.3    &   0.6  \\
\hline
{ BP III}     & 106.6     & 44.8     & 17.7       &6.9     &  2.8  \\
\hline
{ BP IV}      &  75.7     & 32.0     & 12.7       &5.0     &  2.1  \\
\hline
{ BP V}       &  23.8     & 12.2      & 5.6      & 2.5     &  1.1  \\
\hline
 { BP VI}     &  38.4     & 17.7      & 7.3       &3.0     &  1.2  \\
\hline
 { BP VII}    &   0.2     &  0.1      & 0.1       &0.0     &  0.0  \\
\hline
{ BP VIII}    &  13.4     &  6.1      & 2.6       &1.1     &  0.5  \\
\hline
{ BP IX}      & 134.5     & 56.2      &22.0       &8.6      & 3.3  \\
\hline
{ BP X}       & 128.9     & 71.5      &36.0      &17.7      & 8.6  \\
\hline
\end{tabular}
\caption{\footnotesize \it Number of events for $\lum =$ 1 $\ifb$ 
in different invariant mass bins in benchmark scenarios presented
in Table 6. We consider bins with
$M_{\gamma \gamma}>200$ GeV or more at LHC 7 TeV run.}\label{result_7}
\end{center}
\end{table}

We next compute  invariant mass distribution of the di-photons 
from h-decays, the squark-gluino events and the SM 
backgrounds.  
We show in Table \ref{result_7} the total number of events  in 
all 
bins with $M_{\gamma\gamma}\geq$ 200, 300, 400, 500 and 600 GeV for 
different benchmark points for $\lum =$ 1$\ifb$. We have checked 
that the
number of events from the Higgs signal in these bins are 
indeed negligible  but sizable contributions come from the other two
sources.  It is interesting to note that the contribution from  the 
squark-gluino events are statistically significant in some cases.  
For example, the total number of background events
($B$) in all bins with $M_{\gamma\gamma}\geq$ 300
for $\lum =$ 2$\ifb$ is 464. The corresponding number of 
squark-gluino induced  events ($S$) for 
in BP X is 143. Hence $S/ \sqrt{B} =$ 6.64 in this case. 
Similarly for { BP IX} the contribution from squark- gluino production 
is above 5 
$\sigma$ fluctuation of the expected SM background. The number of 
events for BP III and IV are also reasonably large albeit with 
smaller significance at this $\lum$.

\begin{figure}[t]
\centerline{\epsfig{figure=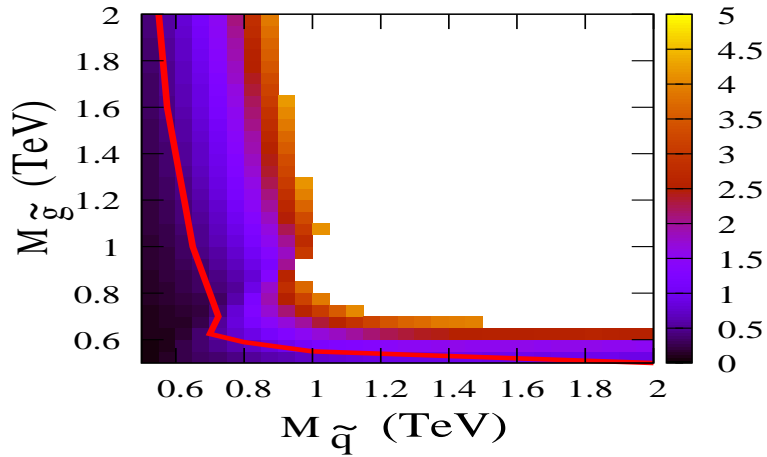,height=6cm,width=10cm,angle=0}}
%\centerline{ \epsfxsize= 6.5 cm\epsfysize=5.0cm  \epsfbox{fig1.eps} }
 \caption{\it The parameter space in a simplified GGMSB model (the 
coloured region) where the distortion in the tail of the $\gamma - 
\gamma$ invariant mass distribution is observable at the ongoing 
LHC experiments with $\lum \leq$5 
$\ifb$ (see text for the details). The parameter space below the red 
line has already been ruled out by the CMS 
collaboration \cite{cmsdiphoton}. 
} \label{fig:reach7}
\end{figure}

\begin{table}[b]
\begin{center}
\begin{tabular}{|c|c|c|c|c|}
\hline
Mh    & $g g  \rightarrow h$ & Vector Boson Fusion  & $W h $, $Z h$, $t \bar{t} h$ &	Br(h $\rightarrow \gamma \gamma$) \\
(GeV) & NLO (pb)             &       NLO (pb)       & NLO (pb)                     &    \\
\hline               
115   &  39.2              &        4.7           &  3.8                         &  0.00208 \\
\hline
\end{tabular}
\caption{\footnotesize \it  Higgs production cross-sections in the SM at 
the LHC 
for $\sqrt{s}$=14 TeV and the branching ratio 
to the di-photon final state(taken from CMS TDR).}\label{Higgs_cross}
\end{center}
\end{table}

Of course such a small integrated luminosity is insufficient for 
Higgs discovery via the di-photon + X channel at the on going LHC 
experiments \cite{moriond}. 
However, it is encouraging to note that even at 
this low $\lum$, BSM physics can show up in the distorted tail 
of the $\gamma-\gamma$ invariant mass distribution. Moreover, if 
observed, this would certainly disfavour the competing SUSY breaking 
models like mSUGRA which do not have natural sources of such di-photons 
. This observation  would also disfavour a two Higgs doublet 
extension of the SM model with  
comparable h- mass and couplings. Such a model  may predict the 
Higgs peak at the right place but not
the appreciable distortion of  the tail.
appreciably.\\

It may also be noted that the contribution of the mGMSB is negligible 
for 
values of $\Lambda$ constrained by the Tevatron data. In contrast, some 
of the non-minimal scenarios contribute significantly. Moreover, for $R=1$, 
the SUSY contributions even in non-minimal scenarios are 
negligible. Thus if a distortion of the tail is 
indeed observed it may indicate a non-minimal model with $R<1$.

We next consider the simplified GGMSB model studied by the CMS 
collaboration \cite{cmsdiphoton}, briefly discussed at the end of
section 3. We fix $\mlspone$ at 50 GeV. The 
di-photons originating from the squark-gluino events are studied 
for different combinations of the common
squark mass $M_{\tilde{q}}$ and the gluino mass $M_{\tilde{g}}$. The 
results are presented in Figure \ref{fig:reach7} . The 
coloured area corresponds to the region where  $S/\sqrt{B}$ as defined 
above is $\geq$ 5 for
an integrated luminosity determined by the colour codes given on the
vertical line next to the figure. In computing the significance we have 
considered the events accumulated in all bins with $M_{\gamma \gamma} >$
300. The parameter space already excluded by the CMS collaboration 
\cite{cmsdiphoton} from
the $\lum =$ 36 $\ifb$ data is shown by the region below the red line. 
Thus hints of 
GGMSB scenarios can be obtained for a fairly large parameter space.

We next repeat the above analysis for experiments at LHC - 14 TeV.
The cross sections in Table \ref{Higgs_cross} are  from the CMS TDR 
\cite{cmstdr}, Table 2.1 for $m_h  =$ 115 GeV at a proton-proton center-of-mass 
energy of $\sqrt{s}$=14 TeV. The NLO cross sections of the dominant SM backgrounds 
are also taken from \cite{cmstdr}.  \\

The irreducible SM backgrounds discussed before and the di-photon events 
stemming from squark-gluino events in different GMSB models are generated 
by PYTHIA. 

\begin{figure}[h]
\centerline{\epsfig{figure=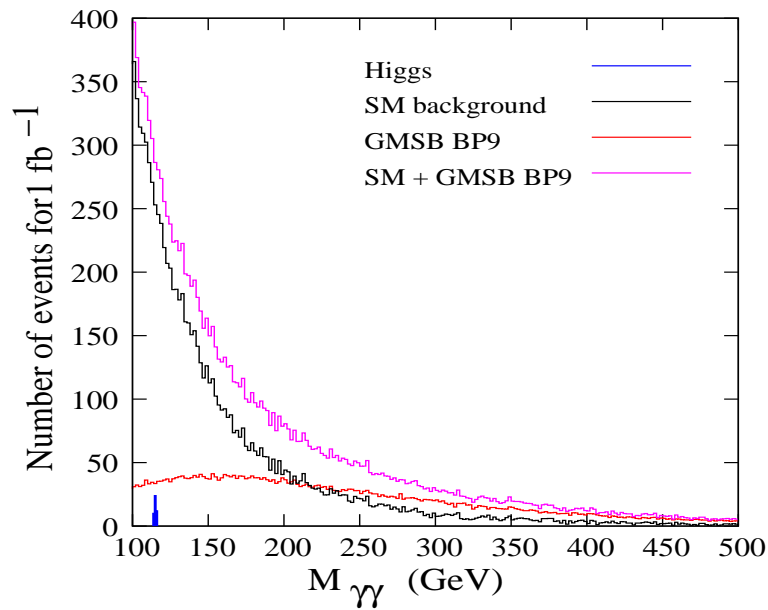,height=8cm,width=10cm,angle=0}}
%\centerline{ \epsfxsize= 6.5 cm\epsfysize=5.0cm  \epsfbox{fig1.eps} }
 \caption{\it The di-photon invariant mass distribution for the SM 
backgrounds, the 
Higgs signal ($m_h=$115 GeV) and squark-gluino events for the GGMSB 
benchmark 
point BP IX at the LHC-14 TeV run with $\lum$ = 1 $\ifb$.  }\label{inv}
\end{figure}

\begin{table}[htbp]\footnotesize
\begin{center}
\begin{tabular}{|c|c|c|c|c|}
\hline
Processes& $M_{\gamma \gamma} > 200$ & $M_{\gamma \gamma} > 400$ & $M_{\gamma \gamma} > 600$& $M_{\gamma \gamma} > 1000$\\
\hline
Born (SM bg)                  &     1275     & 185  &      54  &       8  \\
\hline 
Box  (SM bg)                  &      302     &  15  &       1  &       0  \\
\hline
 BP I                         &       54     &  21  &       7  &       1  \\
\hline
  BP II                       &      214     &  60  &      18  &       2  \\ 
 \hline
 BP III                       &     1876     & 434  &      98  &       7  \\
 \hline 
 BP IV                        &     1519     & 358  &      82  &       6 \\
 \hline
 BP V                         &      547     & 177  &      51  &       5 \\
 \hline
 BP VI                        &      621     & 173  &      42  &       4 \\
 \hline
 BP VII                       &       2      &   1  &       1  &       0 \\
 \hline
  BP VIII                     &      157     &   45 &      12  &       1 \\
 \hline                                     
 BP IX                        &      2580    &  551 &      112 &       7 \\
 \hline
 BP X                         &      1977    &  653 &      199 &      22 \\ 
 \hline
 \end{tabular}
 \caption{\footnotesize  Number of events for $\lum =$ 1 $\ifb$ 
in the invariant mass bins starting from $M_{\gamma \gamma}>200$ GeV at 
the LHC 14 TeV run.}\label{result_14}
 \end{center}
 \end{table}
We present in Figure \ref{inv} the distribution of $M_{\gamma \gamma}$ 
from the decay of a Higgs with $m_h$=115 GeV (the tiny blue
histogram), the SM backgrounds (the black histogram)
and the squark-gluino events from { BP IX} (the red histogram)for  
$\lum =$ 1 $fb^{-1}$ at 14 TeV center-of-mass energy. The combined 
distribution of the SM backgrounds and the SUSY events is also shown (
the purple histogram). It 
is found that the SM background falls off rapidly for higher values of 
the invariant mass. Whereas, the SUSY contribution is fairly constant 
for a significant part of that region. The distribution of events in
different bins is shown in Table \ref{result_14}. 

On the basis of the earlier simulations \cite{cmstdr} one does not 
expect a statistically significant Higgs signal at this tiny $\lum$.
Yet the distortion of the distribution far away from the Higgs peak is 
already worth noting. For example, the number ($S$) of the  
squark-gluino induced events 
for $M_{\gamma\gamma} \geq$ 400 GeV turns out to be 551. The 
corresponding number ($B$) for the SM background is 200. Thus 
$S~/~\sqrt{B}$, as defined above is  39. 
This suggests that the excess of events coming from sparticle 
production, 
if observed near the tail of the distribution, 
can not be dismissed as  mere statistical fluctuation. 
The number of events with $M_{\gamma\gamma} \geq$ 200 GeV or more for other 
benchmark points are given in Table \ref{result_14}.  
\begin{figure}[htbp]
\centerline{\epsfig{figure=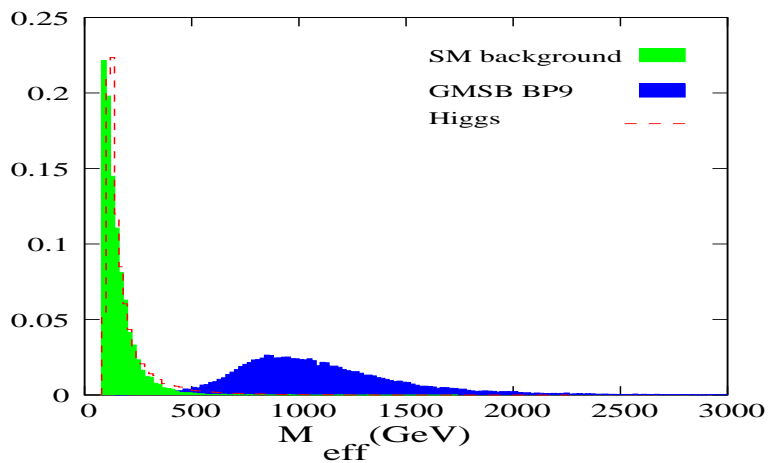,height=6cm,width=10cm,angle=0}}
%\centerline{ \epsfxsize= 6.5 cm\epsfysize=5.0cm  \epsfbox{fig1.eps} }
 \caption{\it Normalized effective mass distribution of the Higgs 
signal (Red dotted line), the GMSB benchmark 
point { BP IX} (blue) and the SM di-photon background (green). }\label{fig:eff}
\end{figure}

The presence of squark gluino events may be revealed by other 
distributions as well. 
We present in Figure \ref{fig:eff} the effective mass distribution of the Higgs signal 
, the SM background and the SUSY events for { BP IX} for $\lum =$ 1 
$\ifb$ at 14 TeV. 
We have defined the effective mass $M_{eff}$ 
as the scalar sum of jet $p_T$'s, lepton $p_T$'s, photon $p_T$'s and 
missing transverse momentum: \begin{equation} M_{eff}= \sum_{jet} p_T + 
\sum_{lepton} p_T + \sum_{photon} p_T + ~~\ptmiss. \end{equation}
The peak of the distribution  at a high $M_{eff}$ 
strongly indicate some BSM physics with new heavy particles and  
disfavour models with an extended Higgs sector but no new 
heavy particles.

\begin{figure}[h]
\centerline{\epsfig{figure=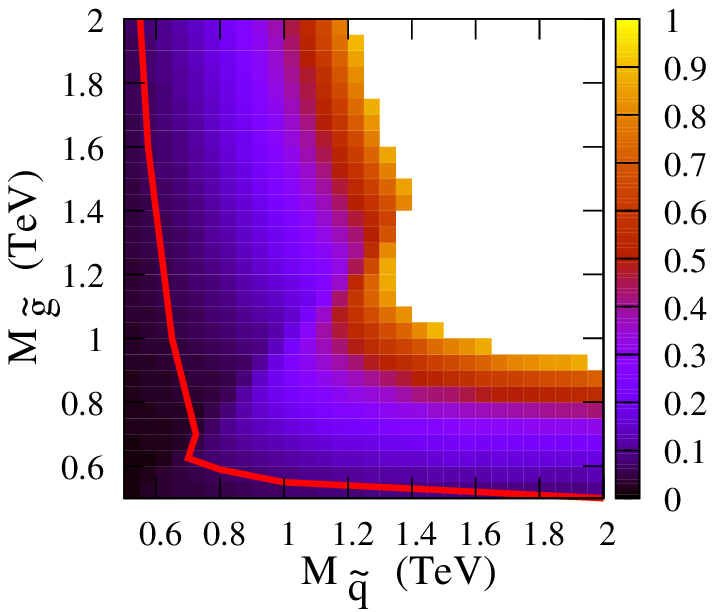,height=6cm,width=10cm,angle=0}}
%\centerline{ \epsfxsize= 6.5 cm\epsfysize=5.0cm  \epsfbox{fig1.eps} }
 \caption{\it Same as Figure \ref{fig:reach7} but for LHC experiments 
at 14 TeV  with 
$\lum \leq 1$ 
$\ifb$.} 
\label{fig:reach14}
\end{figure}
In Fig \ref{fig:reach14} we present our results for the simplified 
model \cite{cmsdiphoton}. The conventions are the same as in Figure
 \ref{fig:reach7} except that we focus on the distortion of the 
$M_{\gamma 
-\gamma}$ distribution for $\lum \leq$ 1 $\ifb$ which will be 
accumulated at a very early stage of experiments at 14 TeV. A large 
region of the parameter space may indicate the distortion looked for. 

A few di-photon events resulting from sparticle production 
in some GGMSB model may enter into the Higgs peak.   
In Table \ref{inv_h} we present the di-photon + X events in the 
neighbourhood of the Higgs peak at 115 for different scenarios for $\lum 
= 1 \ifb $ at 14 TeV. In a bin of $m_h \pm 1.4$, as suggested in 
\cite{atlastdr}, we find the number of genuine Higgs induced events to 
be 43. The 
SUSY contributions to the same region are 33, 24, 47 and 19 for { 
BP III}, { BP IV}, { BP IX} and { BP X} respectively. Thus
the presence of SUSY events is indeed significant.
\begin{table}[htbp]\footnotesize
\begin{center}
\begin{tabular}{|c|c|c|c|c|c|c|c|c|c|}
\hline
Processes&109 &110&111&112&113&114&115&116&117\\
\hline
Born (SM bg)        & 90 &  87 &  83  &  82  &  81  &  85  &   76   &    72   &   67 \\
\hline 
Box  (SM bg)        & 66 & 66  &  60  &  59  &  57  &  54  &   57   &    49   &   50  \\
\hline
{ BP I }      &  0.2 &    0.2 &    0.2  &   0.2  &    0.2  &    0.2  &   0.2  &   0.2  &   0.2 \\
\hline
~{ BP II }  &  1.2 &    1.2 &    1.1  &   1.2  &    1.2  &    1.3  &   1.2  &   1.2  &   1.1 \\
\hline
~~ { BP III }   & 10.6 &   11.4 &   11.9  &  11.7  &   11.1  &   12.2  &  11.8  &  11.5  &  11.3 \\
\hline 
~{ BP  IV }   &  8.1 &    8.1 &    8.7  &   8.1  &    8.1  &    8.8  &   9.0  &   8.4  &   8.8 \\
\hline
{ BP  V}   &  1.8 &    1.9 &    1.9  &   2.1  &    1.9  &    2.0  &   1.6  &   2.0  &   2.1 \\
\hline
~{ BP  VI}   &  2.1 &    2.6 &    2.7  &   2.7  &    2.8  &    2.3  &   2.7  &   3.4  &   2.5 \\
\hline
~~{ BP  VII}   &  0.0 &    0.0 &    0.2  &   0.0  &    0.0  &    0.0  &   0.0  &   0.0  &   0.0 \\
\hline
~~~{ BP  VIII}   &  0.6 &    0.8 &    0.8  &   0.7  &    0.8  &    2.7  &   0.5  &   0.6  &   0.9 \\
\hline               
~{ BP  IX}   & 16.6 &   19.6 &   16.5  &  16.8  &   17.1  &   17.6  &  16.8  &  17.1  &  17.5 \\
\hline 
{ BP  X}   &  6.0 &    6.5 &    6.3  &   6.1  &    6.8  &    6.7  &   6.5  &   6.5  &   6.9 \\
\hline
\end{tabular}
\caption{\footnotesize \it Number of events in the invariant mass bins 
109 
to 117 GeV from different sources after all cuts for $\lum =$ 1 $\ifb$
at the 14 TeV run.
The total Higgs cross section after passing all cuts is around 43 fb 
leading to 43 events at 1 $\ifb$  
distributed in 114-116 GeV bins.} \label{inv_h}
\end{center}
\end{table}

Of course a suitable cut can easily eliminate these unwanted
events. For example, the distribution in Figure \ref{fig:eff} 
suggests that a cut of $M_{eff}<$ 
300 can eliminate the SUSY contributions around the Higgs mass peak, 
while the Higgs signal 
and the SM backgrounds remain practically unaffected. Before the cut, 
however, the non-standard origin of the Higgs boson will be 
revealed by the $M_{eff}$ distribution. The SUSY events around the Higgs peak can
also be eliminated by putting an upper cut on the $\etslash$. However, 
as already discussed the $M_{eff}$ distribution provides a strong hint
for the presence of new heavy particles, whereas large $\etslash$
can come from massless particles and/or jet energy mis-measurements.     

\section{Conclusion}
In this paper we analyze the invariant mass distribution of the 
di-photons to be minutely studied by 
the  dedicated Higgs search experiments in the di-photon + X 
channel at the LHC. In the SM only a 
tiny peak at $m_h$ above the SM 
background is expected. In GMSB models, in contrast, the di-photons
stemming from squark-gluino events may  distort the tail of 
the distribution far away from the Higgs mass peak
revealing the non-standard origin of the Higgs boson. In 
this paper we explore this possibility  both at LHC 7 TeV and 14 TeV
experiments by considering benchmark points in different     
GMSB models - minimal as well as non-minimal (Table 5) 
\cite{GMSB,GMSB1,lowscale,small,metastable,abel,ggmsb}. We 
also consider a simplified model recently employed  by the CMS 
collaboration
to interpret their negative results for SUSY search in the GGMSB models
and illustrate the above point through a detailed parameter space scan.    

We first consider the minimal GMSB \cite{GMSB,GMSB1} model subject to 
the lower bounds on
$m_h$ from LEP and on the lightest neutralino (NLSP) mass
(or equivalently on the scale $\Lambda$)
from Tevatron (see Figures \ref{fig:Mh_tb} and \ref{fig:scan}). The resulting lower bounds on the 
squark-gluino masses are so strong that di-photon events induced by 
sparticle 
production with tiny cross-sections are unlikely to affect 
the h-signal. This has been illustrated with the help of two benchmark 
points (BP I and II, Table \ref{BP_mGMSB} ) in section 3. 

The $D\O{}$ and the CDF  experiments actually 
constrain  quantity $\sigma_{SUSY} \times BR (\lspone \rightarrow 
\gamma~~\tilde{G}) ^2$, where $\sigma_{SUSY}$ is the total sparticle 
production cross section within the kinematic reach of the Tevatron, BR 
($\lspone \rightarrow \gamma~~\tilde{G})$ is the branching ratio 
underlying each single photon event. The bounds on $\Lambda$ or the NLSP
mass in the mGMSB model are derived with the assumption that the above 
BR is one. 

If BR 
$(\lspone \rightarrow \gamma~~\tilde{G})^2$ is significantly smaller 
than 1, the limit on $\sigma_{SUSY}$ and, consequently, on $\Lambda$ or  
the NLSP mass become weaker. This can happen in the mGMSB model (Table 
\ref{branching}) and, more importantly, in its generalizations 
(Table 6). While
extracting $\Lambda_{min}$ from the Tevatron data in different models we 
have taken this possible reduction  into account and have selected our 
benchmark points accordingly (see Table \ref{BP_nonminimal}). 
In some of the models the total squark-gluino cross sections are significantly 
larger than the expected  values in the mGMSB model and the relevant
BRs are not too small either (see Table \ref{cross}). More recently it 
has been 
shown that the distinct features of these phenomenological models 
can  be incorporated into theoretically 
well-motivated models\cite{small,metastable,abel,ggmsb}.         

Our main results are as follows: 
\begin{enumerate}
\item  The di-photon + X events from 
sparticle production may change the shape of  the tail of the 
$\gamma-\gamma$
invariant mass distributions which will be analyzed with utmost care by
the LHC experiments. This distortion was shown to be 
statistically significant in several benchmark scenarios of the 
non-minimal GMSB models for both LHC
7 TeV (Table \ref{result_7}) and 14 TeV 
(Table \ref{result_14}) experiments. 
In both cases the
traces of new physics beyond the Higgs sector may show up at the early 
stages of the experiments with $\lum$ insufficient for the discovery of 
the h boson  in the di-photon + X channel. Similar conclusions follow 
by analyzing a simplified model \cite{cmsdiphoton} 
( Figures \ref{fig:reach7} and \ref{fig:reach14}). 

\item Other observables like the effective mass 
distribution of the di-photon events may also contain hints of new 
physics consisting of heavy particles (Figure \ref{fig:eff}). Unexpected shapes 
of this distribution may disfavour
not only the SM  but i) extensions 
of it with  larger Higgs sectors but no new heavy particles, 
ii) models with other SUSY breaking mechanisms like mSUGRA which do not have 
natural sources of di-photon events from  sparticle production and iii) 
the mGMSB model subject to the constraints from LEP and Tevatron.  

\item
The squark-gluino induced $\gamma-\gamma$ events may contaminate the
peak of the invariant mass distribution at $m_h$ (see Table \ref{inv_h}). To 
improve the purity of the h-signal one may implement cuts on suitable
kinematical variables like an upper cut on the effective mass (see 
Figure \ref{fig:eff}) or an upper cut on the missing energy which eliminate the
sparticle induced events in the Higgs peak.    
\end{enumerate}   

%\begin{figure}[t]
%\centerline{\epsfig{figure=fig6.eps,height=6cm,width=10cm,angle=0}}
%\centerline{ \epsfxsize= 6.5 cm\epsfysize=5.0cm  \epsfbox{fig1.eps} }
% \caption{\it Normalized $p_T$ distribution of first and second  
%photons($p_T$ ordered) for CMS benchmark point (Red lines) and modified 
%CMS benchmark point (blue lines) a with $\sqrt{s}$=7 TeV (see text).
%All distributions are normalized to unity. }
% \label{fig:comparison}
%\end{figure}

\end{document}